\documentclass[%
 reprint,
 amsmath,amssymb,
 aps,
 prd
floatfix,
showkeys
]{revtex4-1}

\usepackage[USenglish]{babel}
\usepackage{slashed}
\usepackage{physics}
\usepackage[normalem]{ulem}
\usepackage{graphicx}% Include figure files
\usepackage{dcolumn}% Align table columns on decimal point
\usepackage{bm}% bold math
\usepackage[usenames]{color}
\usepackage{xcolor}
\usepackage[small]{caption}%\usepackage{hyperref}% 
\usepackage{float}
\usepackage{hyperref}
\usepackage{verbatim}

\newcommand{\be}{\begin{equation}}
\newcommand{\ee}{\end{equation}}
\newcommand{\bea}{\begin{eqnarray}}
\newcommand{\eea}{\end{eqnarray}}
\newcommand{\beas}{\begin{eqnarray*}}
\newcommand{\eeas}{\end{eqnarray*}}

\begin{document}

\title{Damping of density oscillations from bulk viscosity in quark matter}% Force line breaks with \\


\author{Jos\'e Luis Hern\'andez$^{1,2,3}$, Cristina Manuel$^{1,2}$ and Laura Tolos$^{1,2,4}$  }
\affiliation{%
$^1$Institute of Space Sciences (ICE, CSIC), Campus UAB,  Carrer de Can Magrans, 08193 Barcelona, Spain\\
$^2$Institut d'Estudis Espacials de Catalunya (IEEC), 08034 Barcelona, Spain \\
$^3$Facultat de Física, Universitat de Barcelona, Martí i Franquès 1, 08028 Barcelona, Spain.\\
$^4$ Frankfurt Institute for Advanced Studies, Ruth-Moufang-Strasse 1, 60438 Frankfurt am Main, Germany
}

\begin{abstract}

We study the damping of density oscillations in the quark matter phase that might occur in  compact stars. To this end we compute the bulk viscosity and the associated damping time in three-flavor quark matter, considering both nonleptonic and semileptonic electroweak processes. We use two different equations of state of quark matter, more precisely, the MIT bag model and perturbative QCD, including the leading-order corrections in the strong coupling constant. We  analyze the dependence of our results on the density, temperature and  value of strange quark mass in each case.
 We  then find that   the maximum of the bulk viscosity  is in the range of temperature from 0.01 to 0.1 MeV
for frequencies around 1 kHz, while the associated minimal damping times of the density oscillations at those temperatures might be in the range of few to hundreds  milliseconds. Our results suggest that bulk viscous damping might be relevant in the postmerger phase after the collision of two neutron stars if  deconfined  matter is achieved in the process.

\end{abstract}

\keywords{Three-flavor quark matter, Bulk viscosity, Damping time}

\maketitle

%%%%%%%%%%%%%%%%%%%%%%%%%%%%%%%%%%%%%%%%%%%%%%%%%%%%%%%%%%%%%%%%%
\section{Introduction}\label{sec1-CM}
%%%%%%%%%%%%%%%%%%%%%%%%%%%%%%%%%%%%%%%%%%%%%%%%%%%%%%%%%%%%%%%%%

The long-debated possibility that quark matter may be present  within the core of neutron stars, or adopting the form of quark stars, has been extensively explored \cite{Freedman1978PRD}
(see also \cite{Alford:2019oge} for a review and references). While most investigations predominantly focused on assessing the mass-to-radius ratio of the stars, dictated by the equation of state (EOS) linked to the stellar material, recent opportunities for getting insights from neutron star interiors have arisen thanks to the detection of gravitational waves \cite{LIGOScientific:2017vwq,LIGOScientific:2017ync}. In the events of neutron star mergers or in the exploration of diverse stellar oscillation modes generating gravitational radiation \cite{Kokkotas:1999bd,Rezzolla:2003ua,Sieniawska:2019hmd}, nonequilibrium processes unfold. The dynamics of these processes are  influenced by the material's transport coefficients. A comprehensive understanding of the transport coefficients of ultradense matter becomes imperative, as these are determined by the microscopic composition and the dominant interactions of its constituents. The knowledge of the transport coefficients then brings  a connection of the microscopic and macroscopic dynamics of the star.

In this work we will focus on the damping of density oscillations of quark matter, which are relevant in the study of neutron star mergers~\cite{Alford:2017rxf}. We aim to dilucidate whether 
in quark matter dissipative processes might affect the dynamics in neutron star mergers, in the event that a deconfined phase is achieved in the process.  The damping of density oscillations is mainly governed by the bulk viscosity, which is the transport coefficient quantifying the energy dissipation in a compression or rarefaction of matter.

The bulk viscosity of quark matter has been previously studied in \cite{Sawyer:1989uy,Madsen:1992sx,Sad:2007afd,Alford_2006} mainly to determine its effect
on the damping of the so-called r modes on isolated compact stars. The computation of the bulk viscosity of quark matter has also been recently reviewed in  \cite{CruzRojas:2024etx}, improving the form of the EOS of quark matter, both using a perturbative QCD approach and taking into account a nonleptonic electroweak (EW) process, and also using two holographic models. Here we will also compute  the bulk viscosity. We first consider   the MIT bag model to describe quark matter, as this is extensively used in astrophysical settings, but we  also use a QCD perturbative approach. We include both nonleptonic and semileptonic processes in our computation, as the last are relevant in a certain range of  temperatures, as we will show. Unfortunately, at the densities one might expect to find in astrophysical settings, which would hardly exceed 10 times the value of nuclear saturation density $n_0 \approx 0.15 \,{\rm fm}^{-3}$,
the QCD coupling constant $\alpha_s$ is not  so small, and higher-order corrections than those we consider might be needed.
It has been claimed  that the computations of the EOS for quark matter only converge for values of the density  $ \approx 40n_0$ \cite{Kurkela_2010,Gorda:2023usm,Gorda:2023mkk}, values which however are not realistic in astrophysical settings.
That is why one has to invoke some modeling for the EOS of quark matter. It is however instructive to compare those results from those predicted in perturbation theory,  as another possible model for the description of quark matter in compact stars~\cite{Fraga_2001PRD}.

We will further compute the timescale associated to the damping of the density oscillations. This brings relevant information on whether these dissipative processes might be relevant or not, for example, in the inspiral phase of neutron star mergers, or its postmerger dynamics. Attempts to include the effects of bulk viscosity in the numerical modeling of viscous relativistic hydrodynamics valid for neutron stars and neutron star mergers have only been recently initiated \cite{Chabanov:2023abq,Chabanov:2023blf} (see also \cite{CamelioPRDI, CamelioPRDII}).

Our study is complementary to the same study carried out for nuclear matter in Refs.~\cite{Alford_Harris_2019, Alford_2023updated, Alford-Hyperons2021}, or in \cite{Yang:2023ogo}. Also for nuclear matter the processes responsible for the bulk viscosity are mediated by the EW interactions, by either direct or by modified URCA processes.  Also  the value of the bulk viscosity  and the damping density oscillations strongly depends on the EOS used to describe nuclear matter.

This paper is structured as follows. In Sec.~\ref{secII} we present the general framework we use for the computation of the bulk viscosity in the presence of a  periodic disturbance in three-flavor quark matter, and associated to different EW processes. We provide numerical values of the viscosity and corresponding damping times of density oscillations in Sec.~\ref{dampinsection}, using the EOS associated to the MIT bag model in Sec.~\ref{Mitbag}, and also to perturbative QCD in Sec.~\ref{pqcd}. We  present a discussion of our results in Sec.~\ref{secIV}. And, finally, in the Appendix we justify why we ignore temperature effects in the EOS in the temperature range we are considering for the MIT bag model.

We use natural units throughout the article, $\hbar=c= K_B=1$

%%%%%%%%%%%%%%%%%%%%%%%%%%%%%%%%%%%%%%%%%%%%%%%%%%%%%%%%%%%%%%%%%
\section{Bulk viscosity in three-flavor quark matter}\label{secII}
In this section we study the bulk viscosity generated in three-flavor quark matter by nonleptonic and semileptonic weak processes. As a result, we get the bulk viscosity associated to neutrino-free quark matter as a function of temperature $T$, the chemical potentials of the constituent particles of the star, $\mu_i$, and the 
frequency of the oscillation mode, $\omega$. In this article we will only focus on the study of density oscillations, although our results for the
bulk viscosity might be also used for the study of the damping of different stellar oscillation modes. 
In the normal phase and the neutrino-transparent regime, we consider the following equilibration processes:
\begin{eqnarray}
\label{EW-pocesses}
&& u+d \leftrightarrow u+s \ , \\
&& u+e^- \rightarrow d+\nu_e \ , \\
&& d \rightarrow u+e^-+\bar{\nu}_e \ , \\
&& u+e^- \rightarrow s+\nu_e \ , \\
&& s \rightarrow u+e^-+\bar{\nu}_e \ ,
\end{eqnarray}
that involve electrons ($e$) and electronic neutrinos and antineutrinos ($\nu_e$ and $\bar \nu_e$, respectively) as well as up ($u$), down ($d$) and strange ($s$) quarks.

On the one hand, fluctuations around the equilibrium value of the four-vector velocity ($u^{\mu}$) and the particle number density ($n_j$) can be expressed as follows:
\begin{equation}
u^\mu =u^\mu_{0} +\delta u^\mu, \quad n_j=n_{j,0}+\delta n_j,
\end{equation}
such that, in beta equilibrium, we have
\begin{equation}
\partial_{\mu} \left(n_j u^{\mu} \right) =0.
\label{particleconservation}
\end{equation}
Considering the local rest frame (LRF) in the equilibrium state, that is  $u^\mu_0=u^\mu_{\rm LRF}=(1,0,0,0)$ in natural units and neglecting quadratic terms in the deviations, $\mathcal{O}(\delta^2)$, Eq.~\eqref{particleconservation} implies that 
the particle conservation law can be expressed as 
\begin{equation}
\theta n_{j,0} + u^\mu_{\rm LRF} \partial_\mu \delta n_j(t)=0,
\end{equation}
where $\theta=\partial_\mu \delta u^\mu$ is the fluid expansion rate or equivalently
\begin{equation}
\theta n_{j,0} +\frac{\partial}{\partial t} \delta n_j(t)=0.
\label{particleconservationlaw}
\end{equation}
%On the other hand, out-of-beta-equilibrium deviations of the particle number density of the constituent particles are obtained from the following set of equations for the particle density of strange quarks ($n_s$) and the particle density of electrons ($n_e$)
%\begin{eqnarray}
%    &&\frac{\partial}{\partial t} \delta n_s(t) +\theta n_{s,0}=-\lambda_1 \mu_1 -\lambda_2 \mu_2, \\ 
%    &&\frac{\partial}{\partial t} \delta n_e(t) +\theta n_{e,0}=\lambda_2 \mu_2 +\lambda_3 \mu_3,
%\end{eqnarray}
%where $\lambda_1, \, \lambda_2, \, \text{and} \, \lambda_3$ are defined in terms of the equilibration rates of the nonleptonic and semileptonic processes as 
%\begin{eqnarray}
%    \mu_1 \lambda_1 &=& \Gamma_{s+u \rightarrow d+u} - \Gamma_{d+u \rightarrow s+u}, \\
%    \mu_2 \lambda_2 &=& \Gamma_{s \rightarrow u+e+\bar{\nu}_e} - \Gamma_{u+e \rightarrow s+\nu_e}, \\
%    \mu_3 \lambda_3 &=& \Gamma_{d \rightarrow u+e+\bar{\nu}_e} - \Gamma_{u+e \rightarrow d+\nu_e}.
%\end{eqnarray}
On the other hand, out-of-beta-equilibrium deviations generate contributions to the particle density current divergence of the constituent particles. These can be studied in terms of chemical imbalances from beta equilibrium:
\begin{eqnarray}
    && \mu_1 =\mu_s -\mu_d, \\
    && \mu_2 =\mu_s -\mu_e -\mu_u,\\
    && \mu_3 =\mu_d -\mu_e -\mu_u.
\end{eqnarray}
%Note that $\mu_2 -\mu_1=\mu_d -\mu_e -\mu_u=\mu_3 $.

A set of equations for the divergence of the particle density current of strange quarks ($n_su^\mu$) and electrons ($n_e u^\mu$) can be written at linear order in the chemical imbalances as
\begin{eqnarray}
    &&\frac{\partial}{\partial t} \delta n_s(t) +\theta n_{s,0}=-\lambda_1 \mu_1 -\lambda_2 \mu_2, \\ 
    &&\frac{\partial}{\partial t} \delta n_e(t) +\theta n_{e,0}=\lambda_2 \mu_2 +\lambda_3 \mu_3,
\end{eqnarray}
where $\lambda_1, \, \lambda_2, \, \text{and} \, \lambda_3$ are related with the equilibration rates of the nonleptonic and semileptonic processes as follows:
\begin{eqnarray}
    \mu_1 \lambda_1 &=& \Gamma_{s+u \rightarrow d+u} - \Gamma_{d+u \rightarrow s+u}, \\
    \mu_2 \lambda_2 &=& \Gamma_{s \rightarrow u+e+\bar{\nu}_e} - \Gamma_{u+e \rightarrow s+\nu_e}, \\
    \mu_3 \lambda_3 &=& \Gamma_{d \rightarrow u+e+\bar{\nu}_e} - \Gamma_{u+e \rightarrow d+\nu_e}.
\end{eqnarray}

The rates have been computed in several studies at tree level in the limit of massless up and down quarks and also considering that the strange quark mass is considerably smaller than the strange quark chemical potential  \cite{Madsen:1993xx, Heiselberg_1992, Heiselberg:1991px, Koch:1991qh, Iwamoto:1980eb,Iwamoto:1982zz,  Anand2009, schwenzer2012longrange}. For pure massless up and down quarks, perturbative corrections to the quark dispersion law in the strong coupling constant $\alpha_s$ are however needed in order to find a nonvanishing value for $\lambda_3$. One then finds
\begin{eqnarray}    \lambda_1&=& \frac{64}{5\pi^3}G_F^2 \sin^2 \Theta_C \cos^2 \Theta_C \mu_d^5 T^2, \\
    \lambda_2&=&\frac{17}{40 \pi} G_F^2 \sin^2 \Theta_C \mu_s m_s^2 T^4, \\
    \lambda_3&=&\frac{17}{15 \pi^2} G_F^2 \cos^2 \Theta_C \alpha_s \mu_d \mu_u \mu_e T^4,
\end{eqnarray}
where $G_F=1.166\times 10^{-5}\, \text{GeV}^{-2}$ is the Fermi coupling constant, $\Theta_C=13.02^\circ$ is the Cabibbo angle, and $m_s$ is the strange quark mass.

Note that $\mu_2 -\mu_1=\mu_d -\mu_e -\mu_u=\mu_3 $. Then, we obtain that
\begin{equation}
    \frac{\partial}{\partial t} \delta n_e(t) +\theta n_{e,0}=(\lambda_2+\lambda_3) \mu_2 -\lambda_3 \mu_1.
\end{equation}
The oscillating parts of the particle density are taken to be proportional to $e^{i\omega t}$, so that
\begin{equation}
    \frac{\partial}{\partial t}\delta n_j(t)=i\omega \delta n_j(t).
\end{equation}
Thus the equation for the strange quark density can be expressed as follows:
\begin{equation}
    i\omega \delta n_s(t)+\theta n_{s,0}=-\lambda_1 \mu_1 -\lambda_2 \mu_2.
\end{equation}
We can then determine the out-of-beta-equilibrium deviations of the particle number density of electrons and strange quarks considering
\begin{eqnarray}
    &&n_d=2n_B-n_s-n_e, \\
    &&n_u=n_B+n_e. 
\end{eqnarray}
Here we used the charge-neutrality condition 
\begin{equation}
    n_e+\frac{1}{3}n_s +\frac{1}{3}n_d =\frac{2}{3}n_u,
    \label{chargeneutral}
\end{equation}
and the definition of the baryon density
\begin{eqnarray}
    &&n_B \equiv \frac{1}{3} n_u +\frac{1}{3} n_d +\frac{1}{3} n_s. \label{baryonnumber}
\end{eqnarray}
Using the beta-equilibrium and charge-neutrality conditions in the out-of-beta-equilibrium particle number densities, we obtain
\begin{eqnarray}
    &&\delta n_d=2\delta n_B-\delta n_s-\delta n_e, \\
    &&\delta n_u=\delta n_B+\delta n_e.
\end{eqnarray}
%The imbalance of the chemical potentials out-of-beta-equilibrium can be written in terms of deviations of the particle number densities as follows
%\begin{eqnarray}
%    &&\mu_1=A_s \delta n_s -A_d \delta n_d, \\
%    &&\mu_2 =A_s\delta n_s  -A_e \delta n_e -A_u \delta n_u,
%\end{eqnarray}
%where $A_j$ are the susceptibilities of the constituent particles, which can be determined in terms of the following partial derivatives
%\begin{equation}
%A_{ij}=\left(\frac{\partial \mu_i}{\partial n_j} \right),
%\end{equation}
%once we choose an EoS. 
The imbalance of the chemical potentials out of beta equilibrium can be written in terms of deviations of the particle number densities and partial derivatives of the particle's chemical potential with respect to the particle number density:
\begin{equation}
A_{ij}=\left(\frac{\partial \mu_i}{\partial n_j} \right),
\end{equation} 
where $i,j=u,d,s$. From now on we consider only diagonal terms of $A_{ij}$. The off-diagonal terms could also be taken into account depending on the choice of the EOS; we will be back to this issue further on. Thus
\begin{eqnarray}
    &&\mu_1=A_s \delta n_s -A_d \delta n_d, \\
    &&\mu_2 =A_s\delta n_s  -A_e \delta n_e -A_u \delta n_u,
\end{eqnarray}
where $A_j$ are the susceptibilities of the constituent particles and are given by
\begin{equation}
    A_u=A_{uu}, \quad A_d=A_{dd}, \quad A_s=A_{ss},\quad A_e=A_{ee}.
\end{equation}
Employing these relations, we get the conservation equation for the particle number density of strange quarks:
\begin{eqnarray}
    i\omega \delta n_s &=&-\theta n_{s,0}- \lambda_1 [A_s \delta n_s -A_d (2\delta n_B-\delta n_s-\delta n_e)]\nonumber \\
    &-&\lambda_2 [A_s\delta n_s -A_e \delta n_e  -A_u(\delta n_B+\delta n_e)],
    \label{conservationds}
\end{eqnarray}
and a similar expression for the electrons
\begin{eqnarray}
    i\omega \delta n_e &=&-\theta n_{e,0}+(\lambda_2+ \lambda_3)[A_s\delta n_s -A_u (\delta n_B+\delta n_e) \nonumber \\
    &-&A_e \delta n_e]-\lambda_3 [A_s \delta n_s 
    -A_d(2\delta n_B-\delta n_s-\delta n_e)].\nonumber \\
     \label{conservationde}
\end{eqnarray}
Using Eqs.\eqref{conservationds} and \eqref{conservationde} we are able to get an expression for $\delta n_s$. Next, by employing the conservation of the baryon number
\begin{equation}
    \delta n_B =-\frac{\theta}{i\omega} n_{B,0} ,
\end{equation}
we obtain
\begin{eqnarray}
    B\delta n_s &=&-\frac{\theta}{i\omega} \{ i\omega [[i\omega+(\lambda_2 +\lambda_3)A +\lambda_3A_d]n_{s,0}  \nonumber\\
    &+&(\lambda_2 A-\lambda_1 A_d)n_{e,0}]+[A_d(A_u+2A)\lambda_Q \nonumber \\
    &+&i\omega (2\lambda_1 A_d +\lambda_2 A_u)]n_{B,0}   \}, \label{deltans}
\end{eqnarray}
where we define $A\equiv  A_e + A_u$,
\begin{equation}
    \lambda_Q\equiv \lambda_1 \lambda_2+ \lambda_1 \lambda_3+ \lambda_2 \lambda_3 ,\end{equation}
%$\lambda_Q\equiv \lambda_1 \lambda_2+ \lambda_1 \lambda_3+ \lambda_2 \lambda_3 $, 
and
\begin{eqnarray}
    B&\equiv& i\omega[\lambda_1 (A_s +A_d) +\lambda_2 (A+A_s) +\lambda_3 (A+A_d)]\nonumber \\
    &+& \lambda_Q [A(A_s+A_d) + A_d A_s]-\omega^2 .
\end{eqnarray}
And by means of Eqs.~\eqref{conservationde} and \eqref{deltans}, we have
\begin{eqnarray}
    B\delta n_e &=&-\frac{\theta}{i\omega} \{ i\omega [[i\omega +\lambda_1 (A_s+ A_d)+\lambda_2 A_s]n_{e,0}\nonumber \\
    &+&(\lambda_2 A_s -\lambda_3 A_d)n_{s,0}]+ [\lambda_2 A_s(\lambda_2 A_u +2\lambda_1 A_d) \nonumber\\
    &-&[\lambda_2 A_u +\lambda_3 (A_u-2A_d)][i\omega +(\lambda_1 +\lambda_2)A_s]\nonumber \\
    &-&A_d A_u \lambda_Q]n_{B,0}  \}. \label{deltane}
\end{eqnarray}
Out-of-equilibrium deviations of the particle number density, $\delta n_{j}'$ with $j=u,d,s,e$, can be obtained from $\delta n_j$ using the following relation:
\begin{equation}
    \delta n_j=\delta n_{j,0}+\delta n_{j}'
\end{equation}
where $\delta n_{j,0}$ is a fluctuation around beta equilibrium which satisfies
\begin{equation}
    \delta n_{j,0}=-\frac{\theta}{i\omega}n_{j,0}.
\end{equation}
Then, for the electron and the strange quark number density, we get
\begin{eqnarray}
    &&\delta n_e =\delta n_{e,0}+\delta n_e', \\
    &&\delta n_s =\delta n_{s,0}+\delta n_s'.
\end{eqnarray}
Thus
\begin{equation}
    B\delta n_s' =\frac{\theta}{i\omega} [i\omega(\lambda_1 C_1 +\lambda_2 C_2)+(AC_1+A_dC_2)\lambda_Q ]
\end{equation}
and 
\begin{eqnarray}
    B_q\delta n_e' &=&\frac{\theta}{i\omega} \{i\omega[(C_1 -C_2)\lambda_3 -C_2 \lambda_2] \nonumber \\
    &-&[(A_d +A_s)C_2 -A_s C_1]\lambda_Q \},
\end{eqnarray}
where 
\begin{eqnarray}
    C_1 &\equiv& n_{s,0}A_s-n_{d,0}A_d \nonumber \\
    &=& n_{s,0}A_s-(2n_{B,0}-n_{s,0}-n_{e,0})A_d, \\
    C_2 &\equiv& n_{s,0}A_s-n_{u,0}A_u-n_{e,0}A_e \nonumber \\
    &=&n_{s,0}A_s-n_{e,0}A-n_{B,0}A_u.
\end{eqnarray}

Once known the out-of-equilibrium fluctuations of the particle number densities, we can calculate the bulk viscosity. First, we determine the out-of-equilibrium pressure as 
\begin{equation}
    p(n_j (t))=p(n_{j,0}+\delta n_{j,0})+\delta p'(t)=p_0 (t)+\delta p'(t),
\end{equation}
where the nonequilibrium part of the pressure is given by
\begin{equation}
    \Pi =\delta p' =\sum_{j}\left(\frac{\partial p}{\partial n_j} \right)_0 \delta n_j'.
\end{equation}
The nonequilibrium pressure can be expressed in terms of nonequilibrium deviations of the chemical potential by the Gibbs-Duhem equation
\begin{equation}
dp = sd T + \sum_i n_i d \mu_i %add a comma
\end{equation}
where we assume that the thermal equilibrium rate is much larger than the chemical equilibrium rate, thus being the temperature constant (see the note [51] in~\cite{Alford_2021}) so that $dT \approx 0$ which at low temperatures is also equivalent to take baryon density oscillations to be adiabatic.
Thus
\begin{equation}
    c_j \equiv \left(\frac{\partial p}{\partial n_j} \right)_0 =\sum_i n_{i,0} \left(\frac{\partial \mu_i}{\partial n_j} \right)_0 =\sum_i n_{i,0}A_{ij} .
    \label{cj}
\end{equation}
Then the nonequilibrium pressure in quark matter can be expressed as 
\begin{equation}
    \Pi=\sum_j c_j \delta n_j'=c_e\delta n_e'+c_u\delta n_u'+c_d\delta n_d'+c_s\delta n_s'.
\end{equation}
Due to the conservation of the baryon particle number density, its out-of-equilibrium deviation is zero:
\begin{equation}
    \delta n_B'=0,
\end{equation}
and, as a result, we get
\begin{eqnarray}   
    &&\delta n_{u}'=\delta n_e', \\
    &&\delta n_{d}'=-\delta n_s'-\delta n_e'.
\end{eqnarray}
Thus it follows that
\begin{equation}
    \Pi=(c_u-c_d+c_e)\delta n_e' +(c_s-c_d)\delta n_s'.
\end{equation}
According to Eq.~\eqref{cj}, we find that
\begin{eqnarray}
    &&c_s = n_{s,0}A_{ss}, \\
    &&c_u = n_{u,0}A_{uu}, \\
    &&c_d = n_{d,0}A_{dd}, \\
    &&c_e = n_{e,0}A_{ee}. 
\end{eqnarray}
As mentioned before some assumptions about the EOS of three-flavor quark matter with electrons are required in the calculation of the bulk viscosity. In this study we consider diagonal terms for electrons, $A_{ee}$, because they form an ultrarelativistic noninteracting gas. The same applies for the quarks (even if we consider the first correction in $\alpha_s$), so that the particle number density for each flavor does not depend on the chemical potential of the other flavors. Thus,
\begin{eqnarray}
    &&c_s-c_d= C_1, \\
    &&c_s -c_u-c_e=C_2,
\end{eqnarray}
so that 
\begin{equation}
    \Pi=(C_1 -C_2)\delta n_e' +C_1 \delta n_s',
\end{equation}
which can be expressed as 
\begin{eqnarray}
    B\Pi&=&\frac{\theta}{i\omega} \{i\omega[\lambda_1 C_1^2+\lambda_2 C_2^2 +\lambda_3 (C_1-C_2)^2] \nonumber \\
    &+&[A_d C_2^2 +AC_1^2 +A_s(C_1 -C_2)^2] \lambda_Q \}.
    \label{nonequilpressure}
\end{eqnarray}
At first-order hydrodynamics, the bulk viscosity is given by~\cite{Rezzolla2013, Denicol2021, romatschke2019relativistic}
\begin{equation}
    \zeta \equiv -\frac{\text{Re}[\Pi]}{\theta},
\label{bulkdef}
\end{equation}
where the real part of the nonequilibrium pressure can be obtained from Eq.~\eqref{nonequilpressure}. 

As a result the bulk viscosity in three-flavor quark matter in the $\nu$-transparent regime is given by
%The  bulk viscosity in quark matter is given by
%\begin{equation}
%    \zeta \equiv -\frac{\text{Re}[\Pi]}{\theta},
%    \label{bulkdef}
%\end{equation}
%where the real part of the nonequilibrium pressure can be obtained from Eq.~\eqref{nonequilpressure} as
\begin{equation}
    \zeta = \frac{\kappa_1 +\kappa_2 \omega^2}{\kappa_3 +\kappa_4 \omega^2 +\omega^4},
    \label{bulkviscosity}
\end{equation}
where
\begin{eqnarray}
    \kappa_1 &\equiv& \lambda_Q \{ C_1^2[(A+A_d)[A(\lambda_2 +\lambda_3)+A_d \lambda_3] \nonumber \\
    &-&A_d (A\lambda_2 -A_d \lambda_1)] \nonumber \\
    &-&2C_1(C_1 -C_2) [A_d[(A_d +A_s)\lambda_1 +(A+A_d)\lambda_3] \nonumber  \\
    &-&AA_s\lambda_2] \nonumber  \\
    &+&(C_1-C_2)^2[\lambda_1 (A_d+A_s)^2+\lambda_2A_s^2+\lambda_3 A_d^2] \}, \\
    \kappa_2 &\equiv&  \lambda_1 C_1^2+\lambda_2 C_2^2 +\lambda_3(C_1 -C_2)^2 ,\\    
    \kappa_3 &\equiv& \lambda_Q^2 [A(A_s +A_d)+A_d A_s]^2, \\
    \kappa_4 &\equiv& [(A_d +A_s)\lambda_1 +A_s \lambda_2]^2 +2(A\lambda_2 -A_d \lambda_1) \nonumber \\
    &\times& [A_s (\lambda_2 +\lambda_3)-(A_d +A_s)\lambda_3] \nonumber \\
    &+&[A_d \lambda_3 +A(\lambda_2 +\lambda_3)]^2.
\end{eqnarray}
%Eq.~\eqref{bulkviscosity} is equivalent to the expression obtained for the bulk viscosity of $\nu$-free quark matter in Ref.~\cite{Alford_2006}.

%%%%%%%%%%%%%%%%%%%%%%%%%%%%%%%%%%%%%%%%%%%%%%%%%%
%%%%%%%%%%%%%%%%%%%%%%%%%%%%%%%%%%%%%%%%%%%%%%%%%%
\section{Damping time of density oscillations}
\label{dampinsection}
In this section we determine the damping time associated to the bulk viscosity coming from baryon number density oscillations in a medium. Let us assume a small density oscillation described by $\delta n_B= \delta n_{B,0} e^{i \omega t}$, where
$\delta n_{B,0}$ and $\omega$ are the magnitude and frequency of the oscillation, respectively.

The energy density $\epsilon$ stored in a baryonic oscillation with amplitude $\delta n_{B,0}$ can be obtained as
\begin{equation}
    \epsilon = \frac{1}{2} \frac{\partial^2 \varepsilon}{\partial n_B^2} (\delta n_{B,0})^2,
\end{equation}
where the energy density can be computed as 
\begin{equation}
    \varepsilon= \Omega +\sum_i n_i \mu_i
    \label{energydensity}
\end{equation}
and the damping time is defined by
\begin{equation}
\tau_\zeta \equiv \epsilon/(d\epsilon/dt).   
\end{equation}
The energy dissipation time can be related with the bulk viscosity as~\cite{PhysRevD.39.3804,Alford:2017rxf}
\begin{equation}
\frac{d\epsilon}{dt}= \frac{\omega^2 \zeta}{2} \left(\frac{\delta n_{B,0}}{n_{B,0}} \right)^2.
\end{equation}
As a result, the damping time of baryon density oscillations by bulk viscosity is given by 
\begin{equation}
    \tau_\zeta = \frac{n_{B,0}^2}{\omega^2 \zeta} \frac{\partial^2 \varepsilon}{\partial n_B^2} .
    \label{dampingtime}
\end{equation}
Considering an EOS in the limit of zero temperature results in a simplification of the expression leaving the full temperature dependence encoded only in the bulk viscosity. 

%The remaining terms in Eq.~\eqref{dampingtime} are increasing functions of the baryon number density for the MIT bag model and the pQCD EoSs with values between $2.5-4$ GeV from $n_B=3n_0$ up to $6n_0$.  

%%%%%%%%%%%%%%%%%%%%%%%%%%%%%%%%%%%%%%%%%%%%%%%%%%
\subsection{ MIT bag model}
\label{Mitbag}

In this section we determine the values of the chemical potentials and particle number densities of three-flavor quark matter with electrons in the neutrino-transparent regime using as constraints the beta equilibrium and charge neutrality. As a first approximation we can consider the simplest phenomenological bag model at zero temperature. For finite-temperature corrections to the ideal Fermi gas expressions, see the Appendix~\ref{apndx_A}. For this model the thermodynamic potential is given by
\begin{equation}
    \Omega=\sum_{i=e,u,d,s}\Omega_{i}^{(0)} +B_{\rm eff},
    \label{potentialMIT}
\end{equation}
where $B_{\rm eff}$ is the bag constant \footnote{The bulk viscosity and the damping times are independent of the bag constant unless a chemical-potential dependence is included.} and $\Omega^{(0)}_{i}$ is the grand canonical potential of massless electrons and light quarks described as ideal Fermi gases \begin{equation}
    \Omega^{(0)}_e=-\frac{\mu_e^4}{12\pi^2}
    \label{OmegaFermielectrons}
\end{equation}
and
\begin{eqnarray}
    \Omega^{(0)}_f&=&-\frac{N_c}{12\pi^2}  \left[ \mu_f u_f\left(\mu_f^2 -\frac{5}{2}m_f^2 \right) \right. \nonumber \\
    &+& \left. \frac{3}{2}m_f^4 \ln{\left( \frac{\mu_f+u_f}{m_f}\right)} 
    \right],
    \label{OmegaFermiquarks}
\end{eqnarray}
hereafter $f=u,d,s$, $N_c=3$ is the number of colors, $m_f$ denotes the quark mass, and $u_f \equiv \sqrt{\mu_f^2 -m_f^2}$.

From the thermodynamic potential we are able to get the thermodynamic properties of quark matter. Particularly, the number particle density for each particle specie can be calculated by
\begin{equation}
    n_i = -\left(\frac{\partial \Omega}{\partial \mu_i} \right)_{T,V}.
    \label{particlenumberdensity}
\end{equation}
At zero temperature the number densities of  quarks and electrons can be written as follows:
\begin{equation}
    n_f =\frac{N_c}{3\pi^2}(\mu_f^2 -m_f^2)^{3/2}
    \label{densityquarksMIT}
\end{equation}
and
\begin{equation}
    n_e = \frac{1}{3\pi^2}\mu_e^3.
    \label{densityelectronMIT}
\end{equation}
With these expressions at hand, the beta-equilibrium conditions can be expressed as
%in the Fermi surface approximation as
\begin{eqnarray}
    &&\mu_d =\mu_s, \\
    &&\mu_s=\mu_u +\mu_e.
\end{eqnarray}
Using Eq.~\eqref{chargeneutral} (the charge-neutrality condition) and the definition of the baryon number density of Eq.~\eqref{baryonnumber}, we can determine the four chemical potentials and the four number densities ($\mu_d$, $\mu_s$, $\mu_u$, $\mu_e$, $n_d$, $n_s$, $n_u$, and $n_e$)  for a fixed value of the baryon number density.

The susceptibilities can be obtained from Eqs.~\eqref{densityquarksMIT} and \eqref{densityelectronMIT} and are given by
\begin{equation}
    A_{ee}=\frac{\pi^2}{\mu_e^2}
    \label{AeeMIT}
\end{equation}
and
\begin{equation}
    A_{ff}=\frac{\pi^2}{3\mu_f \sqrt{\mu_f^2 -m_f^2}},
\end{equation}
for electrons and quarks, respectively.

For this study we consider different values of the strange quark mass and the baryon number density in terms of the nuclear saturation density, $n_0 = 0.15$ fm$^{-3}$~\cite{PhysRevC.102.044321}. Several studies have determined $m_s$ at a renormalization scale of $2$ GeV using lattice QCD and other techniques~\cite{Workman:2022ynf}, showing that under these conditions $m_s\approx 100$ MeV.  Since in QCD the strange quark mass follows an increasing trend as the strong coupling constant grows, it is intuitive to consider $m_s$ greater than $100$ MeV in a nonperturbative regime. In the MIT bag model, $m_s$ is a degree of freedom; thus to study the $m_s$ dependence of the bulk viscosity, we explore values from $100$ to $300$ MeV. The light quarks are considered massless, as their tiny values  do not have a relevant effect on the quantities of interest. 
%For this study we consider different values of the strange quark mass and the baryon number density in terms of the nuclear saturation density, $n_0=0.15$ $\text{fm}^{-3}$~\cite{PhysRevC.102.044321}. 

Tables~\ref{tab:MITchpparameters} and \ref{tab:MITndparameters} show the values of the chemical potential and number density for quarks and electrons at $m_s=100$ MeV and typical values of the baryon number density in neutron stars.
\begin{table}[H]
\begin{center}
\begin{tabular}{|c|c|c|c|c|}
\hline
$n_{B,0}/n_0$ & $\mu_{u,0}$ & $\mu_{d,0} $ & $\mu_{s,0}$ & $\,\mu_{e,0}\,$ \\ \hline
$3$ & $324.31$ & $331.84$ & $331.84$ & $7.53$ \\ 
$5$ & $384.52$ & $390.91$ & $390.91$ & $6.39$ \\ 
$6$ & $408.61$ & $414.64$ & $414.64$ & $6.03$  \\ \hline
\end{tabular}
\caption{Chemical potentials in MeV for quark matter with electrons imposing charge neutrality and beta equilibrium using the MIT bag model at $m_s=100$ MeV and varying the baryon number density (normalized to nuclear saturation density).}
\label{tab:MITchpparameters}
\end{center}
\end{table} 

\begin{table}[H]
\begin{center}
\begin{tabular}{| c | c | c | c | c | }
\hline
$n_{B,0}/n_0$ & $n_{u,0}/n_0 $ & $n_{d,0}/n_0$ & $n_{s,0}/n_0$ & $n_{e,0}/n_0$  \\ \hline
$3$ & $3.00$ & $3.21$ & $2.79$ & $1.25 \times 10^{-5}$  \\ 
$5$ & $5.00$ & $5.25$ & $4.75$ & $7.66 \times 10^{-6}$   \\ 
$6$ & $6.00 $ & $6.27 $ & $ 5.73$ & $ 6.42\times 10^{-6}$    \\\hline
\end{tabular}
\caption{Particle number densities normalized to the nuclear saturation density imposing charge neutrality and beta equilibrium using the MIT bag model at $m_s=100$ MeV and varying the normalized baryon number density.}
\label{tab:MITndparameters}
\end{center}
\end{table} 

\begin{figure}[H]
\includegraphics[width=0.483\textwidth]{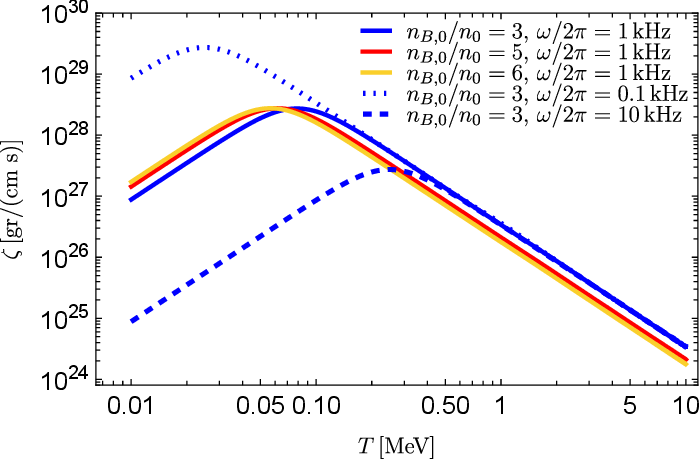}
\centering
\caption{Bulk viscosity of three-flavor quark matter in the neutrino-free regime using the MIT bag model for different normalized baryon number densities and normalized frequencies at $m_s=100$ MeV.}
\label{fig:BVinMITms100}
\end{figure}

\begin{figure}[H]
\includegraphics[width=0.483\textwidth]{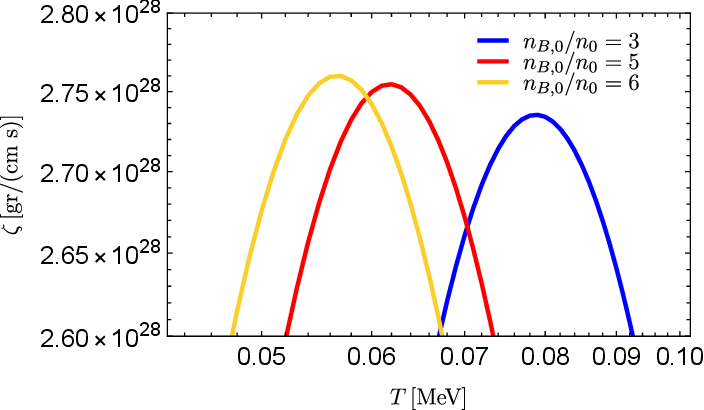}
\centering
\caption{Enlargement of Fig.~\ref{fig:BVinMITms100} to display the behavior of the maximum bulk viscosities at $\omega/2\pi=1$ kHz, $m_s=100$ MeV and for different baryon number densities.}
\label{fig:zoominBVMITms100}
\end{figure}

In Fig.~\ref{fig:BVinMITms100} the bulk viscosity as a function of the temperature is depicted for different frequencies and baryon number densities for a fixed value of $m_s=100$~MeV.  As can be seen,  increasing the baryon number density generates a shift of the maximum of the bulk viscosity to lower temperatures, increasing slightly its value. This is clearly seen in Fig.~\ref{fig:zoominBVMITms100}, where we enlarge the maximum of the bulk viscosities for different densities at $\omega/2 \pi=1$~kHz and $m_s=100$~MeV. The values for the maxima for the different densities at $\omega/2 \pi=1$~kHz and $m_s=100$~MeV are given in Table~\ref{tab:MITresultsfig1and3}. In addition, in Fig.~\ref{fig:BVinMITms100} we consider different values of the angular frequency around $1$ kHz. We observe that the larger the frequency is, the smaller the value of the maximum of the bulk viscosity becomes whereas it moves to larger temperatures.

In order to study the damping times of baryon density oscillations induced by the weak-interaction-driven bulk viscosity we resort to Eq.~\eqref{dampingtime}.
The energy density in the MIT bag model is given explicitly by Eqs.~\eqref{energydensity}, \eqref{potentialMIT}, \eqref{densityquarksMIT}, and \eqref{densityelectronMIT}. 

\begin{figure}[H]
\includegraphics[width=0.483\textwidth]{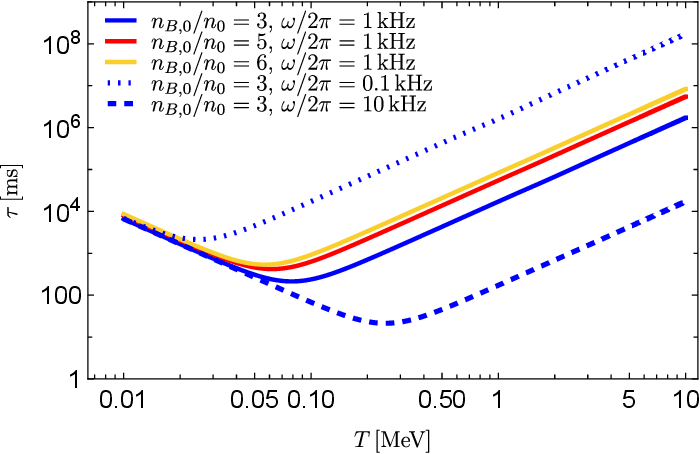}
\centering
\caption{Damping times from density oscillations using the MIT bag model for different normalized baryon number densities and frequencies at $m_s=100$ MeV.}
\label{fig:dampingtimesdiffnB0andomegas}
\end{figure}

%Fig.~\ref{fig:dampingtimesdiffnB0andomegas} displays the damping times associated to the bulk viscosities in Fig.~\ref{fig:BVinMITms100} at $m_s=100$ MeV for different frequencies. We note that the exact values for the minimal damping times can be in found in Table~\ref{tab:MITresultsfig1and3}.
Figure~\ref{fig:dampingtimesdiffnB0andomegas} displays the damping times associated to the bulk viscosities in Fig.~\ref{fig:BVinMITms100}.
%at $m_s=100$ MeV for different frequencies and normalized baryon densities. 
Note that the exact values for the minimal damping times can be  found in Table~\ref{tab:MITresultsfig1and3}. The temperature dependence of the damping time is the same as for the inverse of the bulk viscosity, as this follows from the zero-temperature approximation for the thermodynamic potential. However, all other terms involved in Eq.\eqref{dampingtime} are relevant for determining the exact value of the damping times as a function of the baryon number density. In addition, the $\omega^{-2}$ term modifies significantly the frequency dependence from the inverse of the bulk viscosity.  We also note that at $n_{B,0}/n_0=3$ the damping times seem to be independent of the frequencies considered in the low-temperature regime for approximately $T< 20$ keV. 

\begin{table}[H]
\begin{center}
\begin{tabular}{| c | c | c | c | c| }
\hline
$n_{B,0}/n_0$ & $T_{\rm m}$ & $\zeta_{\rm max}$ & $\tau_{\rm min}$ & $\omega/2\pi$\\ \hline
$3$ & $2.5\times  10^{-2}$ & $2.73\times 10^{29}$ & $2132.42 $ & $0.1$ \\ 
$3$ & $ 2.5\times 10^{-1}$ & $2.73\times 10^{27}$ & $21.32 $  & $10$  \\ 
$3$ & $7.9\times  10^{-2}$ & $2.73\times 10^{28} $ & 
$213.24$  &$1$  \\
$5$ & $6.2\times 10^{-2}$ & $2.75\times 10^{28}$ & $420.14$ &$1$ \\
$6$ & $5.7\times 10^{-2}$ & $2.76\times 10^{28}$ & $535.37$ &$1$ \\\hline
\end{tabular}
\caption{Maximum of the bulk viscosity (in gr cm$^{-1}$ s$^{-1}$) and minimum of the damping times (in milliseconds) for different normalized baryon number densities and frequencies (in kilohertz) at $m_s=100$ MeV according to Figs.~\ref{fig:BVinMITms100} and~\ref{fig:dampingtimesdiffnB0andomegas}. Here $T_{\rm m}$ denotes the temperature in MeV of the maximum (minima) of the bulk viscosity (damping time).}
\label{tab:MITresultsfig1and3}
\end{center}
\end{table} 

\begin{figure}[H]
\includegraphics[width=0.483\textwidth]{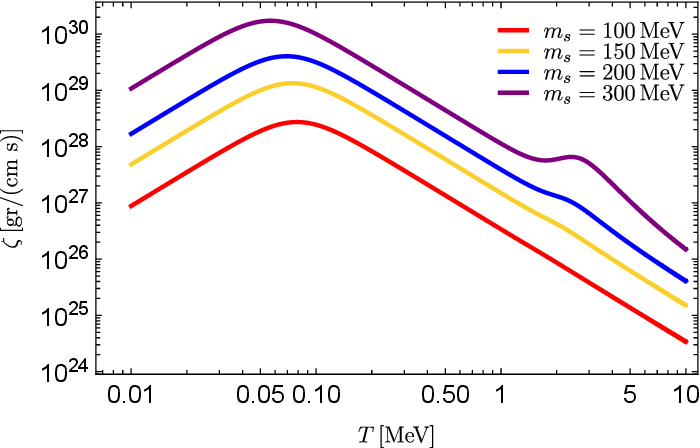}
\centering
\caption{Bulk viscosity of three-flavor quark matter in the neutrino-free regime using the MIT bag model for different values of the strange quark mass at $\omega/2\pi = 1$ kHz and $n_{B,0}/n_0=3$.}
\label{BVvsTMITmodeldiffmass}
\end{figure}

In addition, one can study the effect of the strange quark mass in the bulk viscosity. Figure~\ref{BVvsTMITmodeldiffmass} shows the bulk viscosity as a function of the temperature for different values of the strange quark mass at $\omega/2\pi=1$~kHz and $n_{B,0}/n_0=3$. As the strange quark mass increases, not only is the bulk viscosity larger but also the effect of the semileptonic processes becomes more evident for temperatures of a few MeV. This effect is linked to the fact that $\lambda_2 \propto m_s^2$.
%, particularly those involved in $\lambda_2$

\begin{figure}[H]
\includegraphics[width=0.483\textwidth]{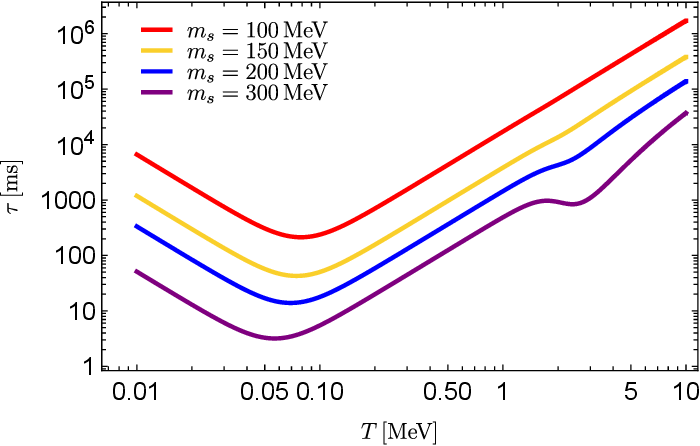}
\centering
\caption{Damping times from density oscillations using the MIT bag model for different values of the strange quark mass at $\omega/2\pi = 1$ kHz and $n_{B,0}/n_0=3$.}
\label{tauvsTMITmodeldiffmass}
\end{figure}

Finally, in Fig.~\ref{tauvsTMITmodeldiffmass} we depict the damping times associated to varying the value of the strange quark mass, corresponding to the viscosities in  Fig.~\ref{BVvsTMITmodeldiffmass}. Increasing the value of the mass of the strange quark has a drastic effect in lowering the damping times below 10 ms. The minimal damping times of density oscillations for  $\omega/2\pi=1$ kHz thus can range from 3 to 200 ms at a given temperature, but this depends strongly on the value of the strange quark mass. The values for the maxima of the bulk viscosity and the minimal damping time are given in Table~\ref{tab:MITresults}.

\begin{table}[H]
\begin{center}
\begin{tabular}{| c | c | c | c | }
\hline
$\,m_s\,$ & $T_{\rm m}$ & $\zeta_{\rm max}$ & $\tau_{\rm min}$ \\ \hline
$100$ & $7.9\times 10^{-2}$ & $2.73\times 10^{28}$ & $213.24$  \\ 
$150$ & $7.5\times 10^{-2}$ & $1.34\times 10^{29}$ & $42.77$   \\ 
$200$ & $6.9\times 10^{-2}$ & $4.05\times 10^{29} $ & 
$13.93$    \\
$300$ & $5.7\times 10^{-2}$ & $1.73\times 10^{30}$ & $3.19$ \\\hline
\end{tabular}
\caption{Maximum of the bulk viscosity (in gr cm$^{-1}$ s$^{-1}$) and minimum of the damping times (in milliseconds) for different masses (in MeV) according  to Figs.~\ref{BVvsTMITmodeldiffmass} and ~\ref{tauvsTMITmodeldiffmass}. Here $T_{\rm m}$ denotes the temperature in MeV of the maximum (minima) of the bulk viscosity (damping time).}
\label{tab:MITresults}
\end{center}
\end{table}

\subsection{Perturbative QCD} 
\label{pqcd}
A similar analysis can be performed for perturbative QCD at high density with a finite mass for the strange quark. As previously stated for the MIT model, a zero-temperature limit for the number particle density and susceptibility of the constituent particles is a good approximation for the temperature region of interest. In perturbative QCD, the thermodynamic potential at finite temperature and chemical potential up to $\mathcal{O}(\alpha_s)$ has been addressed in Ref.~\cite{Fraga_2005}. In this work we consider the zero-temperature limit of this expression which is given by
%The thermodynamic potential per unit volume
\begin{eqnarray}
    \Omega &=& \Omega^{(0)}_e + \sum_{f=u,d,s}\left( \Omega^{(0)}_f +\Omega^{(1)}_f \right),
    \label{potentialquarkmatter}
\end{eqnarray}
where the leading-order terms for massless electrons and nonvanishing quark masses are shown in Eqs.~\eqref{OmegaFermielectrons} and \eqref{OmegaFermiquarks}, and the first-order correction in the $\overline{\text{MS}}$ scheme is given by

\begin{eqnarray}
\Omega^{(1)}_f&=&\frac{\alpha_s (N_c^2-1)}{16\pi^3} \bigg\{3 \left[m_f^2 \ln{\left( \frac{\mu_f+u_f}{m_f}\right)} -\mu_f u_f \right]^2 \nonumber \\ &-&2u_f^4+m_f^2  \left[ \mu_f u_f -m_f^2\ln{\left( \frac{\mu_f+u_f}{m_f}\right)}\right]  \nonumber \\
&\times&\left[6\ln{\left(\frac{\bar{\Lambda}}{m_f} \right)} +4 \right]\bigg\},
\end{eqnarray}
where $\bar{\Lambda}$ is the renormalization scale and $u_f\equiv \sqrt{\mu_f^2 -m_f^2}$ as in the previous section. The thermodynamic potential up to order $\alpha_s$ depends on the renormalization subtraction point explicitly and implicitly through the scale dependence of the strong coupling constant and the mass~\cite{Fraga_2005, Kurkela_2010, Vermaseren_1997}. Considering the massless approximation for the light quarks, the scale dependence of the coupling and the strange quark mass to first order in $\alpha_s$ can be expressed as 
\begin{eqnarray}
    &&\alpha_s(\bar{\Lambda})=\frac{4\pi}{\beta_0 L} \left[1-2\frac{\beta_1}{\beta_0^2} \frac{\ln{(L)}}{L} \right], \label{alphastrong}\\
    &&m_s(\bar{\Lambda})=\hat{m}_s\left(\frac{\alpha_s}{\pi} \right)^{4/9} \left(1+0.895062 \frac{\alpha_s}{\pi} \right), \label{strangemass}
\end{eqnarray}
with $L=2\ln{(\bar{\Lambda}/\Lambda_{\overline{\text{MS}}})}$, the one-loop $\beta$-function coefficient $\beta_0 =11-2N_f/3$, and the two-loop coefficient $\beta_1 =51-19N_f/3$ with $N_f =3$. $\Lambda_{\overline{\text{MS}}}$ and the invariant mass $\hat{m}_s$ can be fixed by requiring $\alpha_s \simeq 0.3$ and $m_s \simeq 100$ MeV at $\bar{\Lambda}=2$ GeV. As a result, one obtains $\Lambda_{\overline{\text{MS}}} \simeq 380$ MeV and $\hat{m}_s =262$ MeV. According to these constraints, the only undetermined parameter is the value of the renormalization scale $\bar{\Lambda}$. 

The particle number density for electrons is given in Eq.~\eqref{densityelectronMIT} and for quarks up to $\mathcal{O}(\alpha_s)$ can be expressed as 
\begin{equation}
    n_f =n_f^{(0)} +n_f^{(1)},
    \label{nfperturbativeqcd}
\end{equation}
where
\begin{equation}
    n_f^{(0)}=\frac{N_c}{3\pi^2}(\mu_f^2-m_f^2)^{3/2}
\end{equation}
and
\begin{eqnarray}
     n_f^{(1)}&=&-\frac{\alpha_s (N_c^2-1)}{4\pi^3} \mu_f u_f^2 \bigg\{1- \frac{3m_f^2}{\mu_f u_f} \ln{\left( \frac{\mu_f+u_f}{m_f}\right)}\nonumber \\
     &+&\frac{m_f^2}{2\mu_f u_f} \left[6\ln\left(\frac{\bar{\Lambda}}{m_f} \right)+4 \right] \bigg\}.
     \label{nf1loopqcd}
\end{eqnarray} 
An alternative to handle Eq.~\eqref{nf1loopqcd} is not to consider the term in brackets that depends on $\bar{\Lambda}$ as done in Ref.~\cite{Sad:2007afd}, which is equivalent to setting $\bar{\Lambda}=\exp(-2/3)m_f$. However, in our approach following this procedure results in a fixed strong coupling constant and strange quark mass according to Eqs.~\eqref{alphastrong} and \eqref{strangemass}. Other alternatives consider $\bar{\Lambda}=2\mu_s$ and $3\mu_s$, which have a relevant impact in the mass-to-radius ratio of compact stars~\cite{Fraga_2001PRD, Fraga_2005}, and $\bar{\Lambda}=2\sum_f \mu_f/N_f$ as in Ref.~\cite{Kurkela_2010}. 

In this study, we proceed setting $\bar{\Lambda}=2\mu_s$ and implement the beta-equilibrium and the charge-neutrality conditions. This procedure differs from the study in Ref.~\cite{Sad:2007afd}, where the bulk viscosity is computed for two different sets of parameters ($n_{B}/n_0 = 5$, $m_s=300$ MeV, and $\alpha_s=0.2$ and  $n_{B}/n_0 = 10$, $m_s=140$ MeV, and $\alpha_s=0.1$). Accordingly, our approach is an alternative to incorporate the trend of the strange quark mass and the strong coupling constant as a function of the baryon number density (see Table~\ref{tab:chpparameters}). As can be noted, the mass and the coupling decrease as the baryon number density increases consistently with the constraint $m_s \simeq 100$ MeV and $\alpha_s\simeq 0.3$ at $\bar{\Lambda}=2$ GeV.

With these expressions at hand, we solve the beta-equilibrium and charge-neutrality conditions for different values of the baryon number density. The light quarks are considered massless. Tables~\ref{tab:chpparameters} and \ref{tab:ndparameters} list the values of the chemical potential, the strange quark mass, the strong coupling constant, and the number density of the constituent particles in three-flavor quark matter with electrons for different baryon number density.
\begin{table}[H]
\begin{center}
\begin{tabular}{| c | c | c | c | c | c | c | }
\hline
$n_{B,0}/n_0$ & $\mu_{u,0} $ & $\mu_{d,0} $ & $\mu_{s,0}$ & $\mu_{e,0} $ & $m_s$ & $\alpha_s$ \\ \hline
$6$ & $470.47$ & $489.18$ & $489.18$ & $18.71$ & $ 138.46$ & $0.54$\\
$10$ & $544.80$ & $557.26$ & $557.26$ & $12.46$ & $ 127.12$ & $0.47$ \\
$20$ & $671.16$ & $678.75$ & $678.75$ & $7.58$ & $ 115.06$ & $0.39$ \\
$40$ & $832.88$ & $837.75$ & $837.75$ & $4.87$ & $ 106.01$ & $0.33$ \\ \hline
\end{tabular}
\caption{Input parameters for the bulk viscosity with pQCD at different normalized baryon number densities imposing beta equilibrium and electric charge neutrality: chemical potentials in MeV, the strange quark mass in MeV and the strong coupling constant.}
\label{tab:chpparameters}
\end{center}
\end{table} 

\begin{table}[H]
\begin{center}
\begin{tabular}{| c | c | c | c | c | }
\hline
$n_{B,0}/n_0$ & $n_{u,0}/n_0 $ & $n_{d,0}/n_0$ & $n_{s,0}/n_0$ & $n_{e,0}/n_0$  \\ \hline
$6$ & $6.00$ & $6.74$ & $5.25$ & $1.92\times 10^{-4}$ \\
$10$ & $10.00$ & $10.70$ & $9.30$ & $5.67\times 10^{-5}$ \\
$20$ & $20.00$ & $20.69$ & $19.31$ & $1.28\times 10^{-5}$ \\
$40$ & $40.00$ & $40.71$ & $39.29$ & $3.39\times 10^{-6}$ \\ \hline
\end{tabular}
\caption{Normalized particle number densities with pQCD
at different normalized baryon number densities imposing beta equilibrium and electric charge neutrality.}
\label{tab:ndparameters}
\end{center}
\end{table} 

The susceptibilities are given by Eq.~\eqref{AeeMIT} and
\begin{eqnarray}
    A_{ff}^{-1}&=&\frac{N_c}{\pi^2}\mu_f \sqrt{\mu_f^2 -m_f^2}-\frac{\alpha_s (N_c-1)^2}{4\pi^3}\bigg\{3\mu_f^2 -4m_f^2 \nonumber \\
    &+& \frac{m_f^2 \mu_f}{u_f}\left[2-3\ln\left(\frac{\mu_f +u_f}{m_f} \right) +3\ln\left(\frac{\bar{\Lambda}}{m_f}\right)  \right] \bigg\}, \nonumber \\
\end{eqnarray}
for electrons and quarks, respectively.

\begin{figure}[H]
\includegraphics[width=0.483\textwidth]{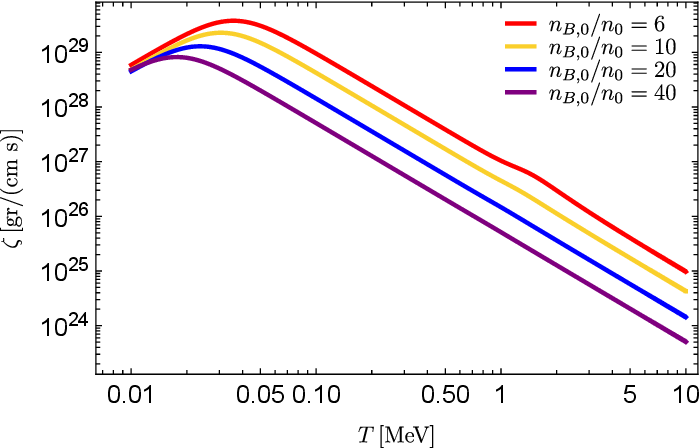}
\centering
\caption{Bulk viscosity of three-flavor quark matter with electrons using perturbative QCD for different normalized baryon number densities at $\omega/2\pi=1$ kHz and and $\bar{\Lambda}=2\mu_s$.}
\label{fig:BVpQCD}
\end{figure}

In Fig.~\ref{fig:BVpQCD} we plot the bulk viscosity as a function of the temperature at $\omega/2\pi =1$ kHz and $\bar{\Lambda}=2\mu_s$ for different baryon number densities. 
% In the region of typical densities of neutron stars the trend of the bulk viscosity is similar to the results in Ref.~\cite{Sad:2007afd}. 
%The maximum of the bulk viscosity lies in the low-temperature regime and a slight increase of the bulk viscosity due to the semileptonic processes at intermediate values of the temperature and that it is more prominent for lower values of the baryon number density.

 %For typical densities of neutron stars the trend of the bulk viscosity is similar to the results in Ref.~\cite{Sad:2007afd}.

Our results seem to qualitatively agree with those recently presented in \cite{CruzRojas:2024etx}, valid for densities $40n_0$, within perturbative QCD, even if in that reference higher-order corrections to the EOS were included, and only the nonleptonic process $u+d \leftrightarrow u+s$ was considered. 
We have checked that the value of the maximum value of the bulk viscosity as well as its location as a function of the temperature qualitatively agree, when computed at the same order of accuracy. In Ref.~\cite{CruzRojas:2024etx} higher-order perturbative corrections are included, changing slightly the position and the value of the maximum of the bulk viscosity. Further deviations with~\cite{CruzRojas:2024etx} are due to the fact that semileptonic processes become relevant in a temperature region around $1-2$ MeV and also give rise to a secondary peak in the bulk viscosity that is most remarkable as the strange quark mass increases and the baryon number density decreases.

%We however find some discrepancies with \cite{CruzRojas:2024etx} that might be due to the fact  that in this reference  the semileptonic processes were discarded.

In order to check the relevance of the semileptonic process we consider a similar approach to Ref.~\cite{Sad:2007afd} computing the
 bulk viscosity generated only by nonleptonic weak processes, $\zeta_{\rm non}$. This can be obtained from the general expression in Eq.\eqref{bulkviscosity} setting $\lambda_2, \lambda_3 \to 0$ and it is given by
\begin{equation}
    \zeta_{\rm non}=\frac{\lambda_1 C_1^2}{(A_d+A_s)^2\lambda_1^2 +\omega^2}.
\end{equation}

\begin{figure}[H]
\includegraphics[width=0.483\textwidth]{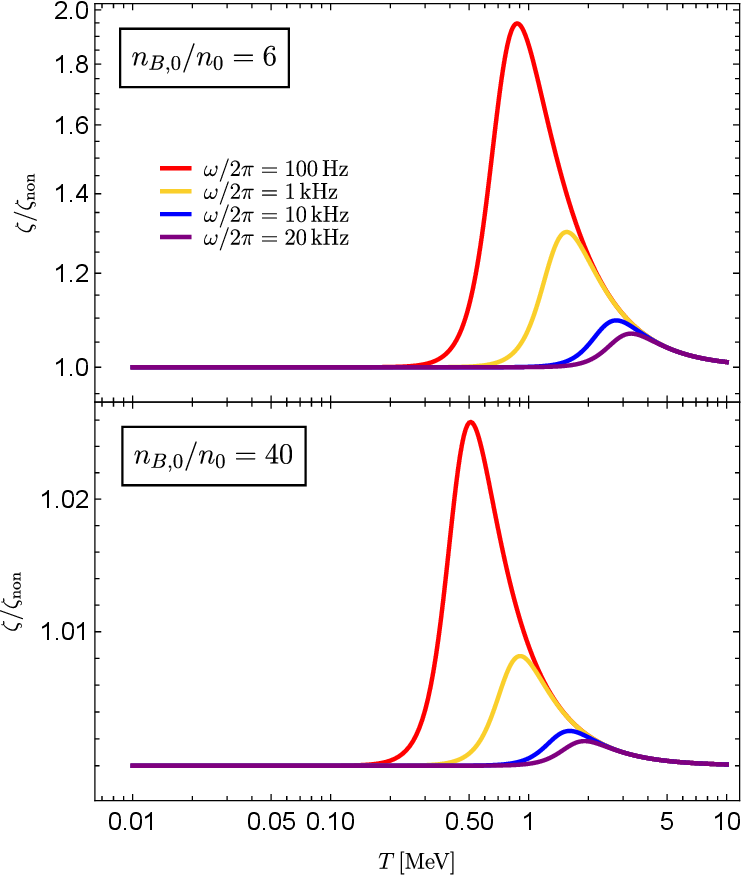}
\centering
\caption{Ratio $\zeta/\zeta_{\rm non}$ as a function of the temperature using perturbative QCD for different frequencies at $n_{B,0}/n_0=6$ and $40$. Note that the y axis is different for each case.}
\label{fig:compare-non-lep}
\end{figure}

We  plot the ratio of the full bulk viscosity with that arising only from the nonleptonic processes 
in Fig.~\ref{fig:compare-non-lep}. We conclude that
there is certain range of temperatures, in the region from 0.1 to 2 MeV, where neglecting the semileptonic processes is not a good approximation. We note that in Ref.~\cite{Sad:2007afd} it has been claimed that the regime where the semileptonic processes might be dropped depends on the value of the frequency at a given value of the density and temperature. 

\begin{figure}[H]
\includegraphics[width=0.483\textwidth]{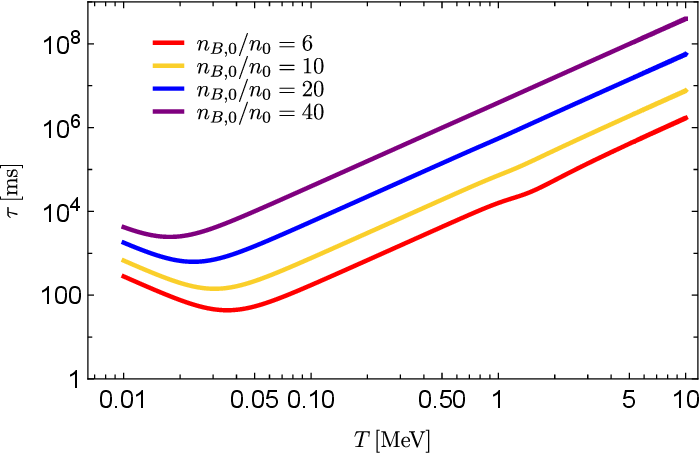}
\centering
\caption{Damping times from density oscillations using perturbative QCD for different baryon number densities at $\omega/2\pi=1$ kHz and $\bar{\Lambda}=2\mu_s$.}
\label{fig:dampingtimespQCD}
\end{figure}

Lastly, Fig.~\ref{fig:dampingtimespQCD} shows the damping times for different baryon densities at $\omega/2\pi=1$ kHz and $\bar{\Lambda}=2\mu_s$. The values for the maxima of the bulk viscosity and the minimal damping time for different densities are given in Table~\ref{tab:pQCDresults}.
As stated before, the strange quark mass decreases as the baryon number density gets larger. We see that the maximum bulk viscosity decreases when the baryon density increases. 
The damping time curves exhibit the same trend as in Fig.~\ref{tauvsTMITmodeldiffmass}. At $n_{B,0}/n_0=6$ we have $m_{s}\approx138$ MeV and for higher values of the baryon number density up to $n_{B,0}/n_0=40 $ we get $m_{s}\approx 106$ MeV.

Note that the maximum bulk viscosity and shortest damping times exhibit the opposite  behavior when increasing the density in the MIT bag model; see Fig.~\ref{fig:zoominBVMITms100}. This is linked to the fact that there  the strange quark mass is a fixed parameter, which does not change with the density. It should be possible to improve this feature in the modeling of the EOS, but we leave it for future work.

\begin{table}[H]
\begin{center}
\begin{tabular}{| c | c | c | c | }
\hline
$n_{B,0}/n_0$ & $T_{\rm m}$ & $\zeta_{\rm max}$ & $\tau_{\rm min}$ \\ \hline
$6$ & $3.6\times 10^{-2}$ & $3.78\times 10^{29}$ & $43.65$  \\ 
$10$ & $3.1\times 10^{-2}$ & $2.28\times 10^{29}$ & $142.12 $   \\ 
$20$ & $2.4\times 10^{-2}$ & $1.29\times 10^{29} $ & 
$626.32$    \\
$40$ & $1.8\times 10^{-2}$ & $8.20\times 10^{28}$ & $2459.50$ \\\hline
\end{tabular}
\caption{Maximum of the bulk viscosity (in gr cm$^{-1}$ s$^{-1}$) and minimum of the damping times (in milliseconds) for different baryon number density according to Figs.~\ref{fig:BVpQCD} and ~\ref{fig:dampingtimespQCD}. Here $T_{\rm m}$ denotes the temperature in MeV of the maximum (minima) of the bulk viscosity (damping time).}
\label{tab:pQCDresults}
\end{center}
\end{table} 

%\begin{figure}[H]
%\includegraphics{comparison.eps}
%\centering
%\caption{Bulk viscosity as a function of the temperature at $\omega/2\pi=1$ kHz and $n_{B,0}/n_0=40$. Plot to compare with Rojas et al.}
%\end{figure}

%%%%%%%%%%%%%%%%%%%%%%%%%%%%%%%%%%%%%%%%%%%%%%%%%%%%%%%%%%%%%%%%%
\section{Outlook}\label{secIV}

We have studied the bulk viscosity and the damping time of density oscillations of quark matter, using different  EOSs, and exploring their dependence on the baryon density, temperature and value of the strange quark mass.
At the densities that could be attained in neutron stars we have considered the MIT bag model and checked that the value of the bulk viscosity changes significatively with the value of the strange quark mass. We have also used an EOS  extracted from QCD at order $\alpha_s$. We have included all the relevant electroweak  processes that equilibrate quark matter after a disturbance of the density and checked in which temperature regime the nonleptonic process is dominant.

While we see that the numerical value of the bulk viscosity of quark matter  depends on the form of EOS, on the value of the strange quark mass, and on the form of the quark dispersion law,
one might see some general features from our results. In particular, we find that
the maximum value of the bulk viscosity, producing the shortest damping times of the density oscillations (in the order of the few to several hundreds of milliseconds, depending on values of the density and the strange quark mass), occurs at temperatures in the range from 0.01  to 0.1  MeV; the precise value depends on the  EOS describing  quark matter.
The bulk viscosity of nuclear matter, which also highly depends on the corresponding modeled EOS of nuclear matter, seems to have its maximum at much higher values of the temperature, in the order of few MeV ~\cite{Alford_Harris_2019, Alford_2023updated}. Then one can clearly conclude the strongest damping of density oscillations occur
in different temperature regimes in quark or nuclear matter, while these different phases occur at different densities.

Our results might be of interest so as to assess whether the effect of the bulk viscosity should be included or not in numerical simulations of mergers of neutron stars. Several such numerical studies mention the possibility of reaching to a deconfined quark matter phase  \cite{Bauswein:2018bma,Most:2018eaw,Weih:2019xvw,Blacker:2020nlq,Prakash:2021wpz}. As the timescales associated to the initial stages of the merger are of the order of few milliseconds,  
unless  there are regions in the stars where the reached temperatures   are in the range of 0.01 MeV, the effect of the bulk viscosity  in the quark matter phase would be unnoticeable. The effect might be more pronounced in the postmerger phase, as it seems also to be the case if one assumes only the presence of nuclear matter; see \cite{Chabanov:2023blf}. However, the effect in both cases depends on the temperatures attained in the postmerger object.

We have not considered the possibility of Cooper pairing of quarks in this article. In the so-called color flavor locked~\cite{Alford:1998mk} phase and much below the superconducting transition the bulk viscosity is dominated by the interaction of the superfluid phonons \cite{Manuel:2007pz} and the kaons~\cite{Alford:2007rw} and it was computed in \cite{Mannarelli:2009ia,Bierkandt:2011zp}. A further study of how density oscillations are damped would be required, but from the results found in \cite{Mannarelli:2009ia,Bierkandt:2011zp} one might predict that damping times would be  longer than in the normal phase.

The effect of the bulk viscosity of quark matter might be also relevant in the study of the damping of the different oscillation modes of  isolated compact stars. We will address them in a different publication.

\section*{Acknowledgments}
Recently, Ref.~\cite{CruzRojas:2024etx} appeared in the arXiv, which has a clear overlap with part of the content of this work. 
We acknowledge support from the program Unidad de Excelencia María de Maeztu CEX2020-001058-M, from Projects No. PID2019-110165GB-I00 and No. PID2022-139427NB-I00 financed by the Spanish MCIN/AEI/10.13039/501100011033/FEDER, UE (FSE+),  as well as from the Generalitat de Catalunya under Contract No. 2021 SGR 171,  by  the EU STRONG-2020 project, under the program  H2020-INFRAIA-2018-1 Grant Agreement No. 824093, and by the CRC-TR 211 \lq \lq Strong-interaction matter under extreme conditions" Project No. 315477589-TRR 211.

\appendix

\section{MIT bag model at finite temperature}
\label{apndx_A}
A typical approach to determine the chemical potentials and susceptibilities of quark matter with electrons inside neutron stars appeals to a zero-temperature limit of the thermodynamic potential. In this appendix we study the temperature dependence of the ideal Fermi gas expressions for the particle density and the susceptibilities to infer its relevance in the temperature region of up to $10$ MeV.

The leading order of the thermodynamic potential for this model at finite temperature is given by~\cite{Wen_2005, Lopes_2021}
\begin{eqnarray}
    \Omega_{i}^{(0)}&=&-\frac{\gamma_i T}{2\pi^2} \int_0^\infty k^2 dk \, \Bigg\{ \ln\left[1+\exp \left(-\frac{ E_{i,k}-\mu_i}{T} \right)\right]  \nonumber \\
    &+&\ln\left[1+\exp\left(-\frac{ E_{i,k}+\mu_i}{T} \right) \right] \Bigg\},
    \label{PotentialidealgasquarksfinT}
\end{eqnarray}    
with $E_{i,k}=\sqrt{k^2 +m_i^2}$ and $\gamma_i$  the degeneracy factor, for electrons, $\gamma_e=2$, accounts their spin degrees of freedom and for quarks, $\gamma_f=2N_c$, considers the spin and color degrees of freedom. For electrons their mass, $m_e=0.511$ MeV, is small compared to the strange quark mass, resorting to the massless approximation. Then, the integration in the thermodynamic potential of an ideal relativistic Fermi gas can be carried out to give
\begin{equation}
    \Omega_e^{(0)}= -\frac{1}{12} \left(\frac{\mu_e^4}{\pi^2} +2\mu_e^2 T^2 +\frac{7}{15}\pi^2T^4 \right).
     \label{PotentialidealgaselectronsfinT}
\end{equation}
Using Eqs.~\eqref{PotentialidealgasquarksfinT} and \eqref{PotentialidealgaselectronsfinT}, the finite-temperature expressions for the number densities are given by
\begin{eqnarray}
    n_f &=&\frac{N_c}{\pi^2} \int k^2 dk \Bigg\{\frac{1}{1+\exp[(E_{f,k} - \mu_f)/T]} \nonumber \\
     &-&\frac{1}{1+\exp[(E_{f,k} + \mu_f)/T]}\Bigg\}
     \label{nffinT}
 \end{eqnarray}
and 
\begin{equation}
 n_e =\frac{ \mu_e }{3} \left( T^2 +\frac{\mu_e^2}{\pi^2} \right),
 \label{nefinT}
\end{equation}
for quarks and electrons, respectively.

Using Eqs.~\eqref{nffinT} and \eqref{nefinT}
we compute the 
chemical potentials of quark matter with electrons imposing the charge-neutrality and beta-equilibrium conditions at different temperatures.

\begin{figure}[H]
\includegraphics[width=0.483\textwidth]{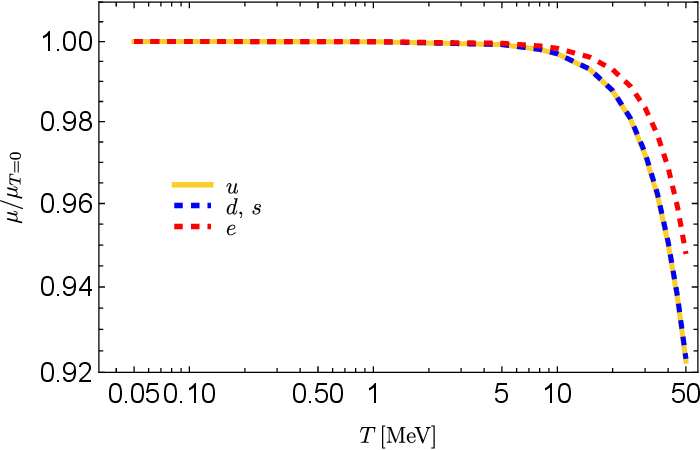}
\centering
\caption{The chemical potentials of the different particle species $\mu_i$ obtained by imposing beta equilibrium and electric charge neutrality normalized by the same value obtained at zero temperature, $\mu_{T=0}$. Additionally, we set $n_{B,0}/n_0=3$ and $m_s=100$ MeV. }
\label{deviationchpTfin}
\end{figure}

In Fig.~\ref{deviationchpTfin} we show the chemical potentials at finite temperature  normalized by its value at zero temperature as well as at $n_{B,0}/n_0=3$ and $m_s=100$ MeV. Thermal effects reduce the chemical potential of the constituent particles in these conditions. For temperatures below $10$ MeV the deviations compared to the value at zero temperature are up to $0.3\%$ for light quarks and up to $0.2\%$ for electrons, while for temperatures up to $50$ MeV these deviations can be up to $8\%$ for light quarks and $\sim 5\%$ for electrons.

To sum up, our predictions for the bulk viscosity values would not be significantly affected by these thermal corrections for temperatures up to $10$ MeV. For higher values of temperatures these effects may be relevant. Also, at these temperatures neutrinos might get trapped in the medium and have to be taken into account in the beta-equilibrium conditions, thus changing the bulk viscosity and associated damping time~\cite{Pal:2011ve}.

%\section{Damping time of adiabatic baryon density oscillations}
%In this appendix  for the purposes of comparing the damping of density oscillations in neutron stars by bulk viscosity with our study in quark matter, we study the behavior of an analogue to the isothermal incompressibility of nuclear matter as a function of the baryon number density at zero temperature. According to the approaches we follow, the temperature-dependence of our results for the damping times are completely encoded in the bulk viscosity so that we expect that 
%\begin{equation}
%    \tau_{\rm min}= \frac{n_{B,0}^2}{\omega^2 \zeta_{\rm max}} \left(\frac{\partial^2 \varepsilon}{\partial n_B^2}\right)_0,
%\end{equation}
\label{apndx_B}
%As stated before, we fix a value of the baryon number density so that the only thermal-dependence that may remain in the damping time expression 

%The isothermal incompressibility of nuclear matter is given by~\cite{particles3020034}
%\begin{equation}
%    K=9n_B\frac{\partial^2 \varepsilon}{\partial n_B^2},
%\end{equation}

%\begin{figure}[H]
%\includegraphics{AnalogueIncompressibility.eps}
%\centering
%\caption{Analogue to the isothermal incompressibility in three-flavor quark matter with electrons imposing beta equilibrium and charge neutrality as a function of the baryon number density. }
%\label{fig:incompressibility}
%\end{figure}

%The isothermal incompressibility of nuclear matter is given by~\cite{particles3020034}
%\begin{equation}
%    K=9n_B\frac{\partial^2 \varepsilon}{\partial n_B^2},
%\end{equation}

%%%%%%%%%%%%%%%%%%%%%%%%%%%%%%%%%%%%%%%%%%%%%%%%%%%%%%%%%%%%%%%%%
\newpage
\bibliography{bibliography}

\end{document}